# Gene Ontology: Pitfalls, Biases, Remedies

Pascale Gaudet[1,2] and Christophe Dessimoz[3,4]

[1] CALIPHO group, SIB Swiss Institute of Bioinformatics, 1211 Geneva, Switzerland
[2] Department of Human Protein Sciences, Faculty of Medicine, University of Geneva, 1211 Geneva, Switzerland
[3] Dept. of Ecology and Evolution & Center for integrative Genomics, University of Lausanne, Biophore, 1015 Lausanne, Switzerland
[4] Dept. of Computer Science & Dept. of Genetics, Evolution and Environment, University College London, Gower Street, London WC1E 6BT, UK
*Correspondence: Pascale.Gaudet@isb-sib.ch and Christophe.Dessimoz@unil.ch*

## Abstract

The Gene Ontology (GO) is a formidable resource but there are several considerations about it that are essential to understand the data and interpret it correctly. The GO is sufficiently simple that it can be used without deep understanding of its structure or how it is developed, which is both a strength and a weakness. In this chapter, we discuss some common misinterpretations of the ontology and the annotations. A better understanding of the pitfalls and the biases in the GO should help users make the most of this very rich resource. We also review some of the misconceptions and misleading assumptions commonly made about GO, including the effect of data incompleteness, the importance of annotation qualifiers, and the transitivity or lack thereof associated with different ontology relations. We also discuss several biases that can confound aggregate analyses such as gene enrichment analyses. For each of these pitfalls and biases, we suggest remedies and best practices.

## 1. Introduction

As we have seen in previous chapters (e.g. ref to Hastings, ref to Bauer, ref to Pesquita), by providing a large amount of structured information, the Gene Ontology (GO) greatly facilitates large-scale analyses and data mining. A very common type of analysis entails comparing sets of genes in terms of their functional annotations, for instance to identify functions that are enriched or depleted in particular subsets of genes (Chap. X) or to assess whether particular aspects of gene function might be associated with other aspects of genes, such as sequence divergence or regulatory networks.

Despite conscious efforts to keep GO data as normalised as possible, it is heterogeneous in many respects—to a large extent simply because the body of knowledge underlying the GO is itself very heterogeneous. This can introduce considerable biases when the data is used in other analysis, an effect that is magnified in large-scale comparisons.

Statisticians and epidemiologists make a clear distinction between *experimental data*—data from a controlled experiment, designed such that the case and control groups are as identical as possible in all respects other than a factor of interest—and *observational data*—data readily available, but with the potential presence of unknown or unmeasured factors that may confound the analysis. GO annotations clearly falls into the second category. Therefore, testing and controlling for potential confounders is of paramount importance.

Before we go through some of the key biases and known potential confounders, let us consider Simpson's Paradox, which provides a stark illustration of the perils of data aggregation.

## 1.1 Simpson's Paradox: the perils of data aggregation

Simpson's paradox is the counterintuitive observation that a statistical analysis of aggregated data (combining multiple individual datasets) can lead to dramatically different conclusions from analyses of each dataset taken individually, i.e. that the whole appears to disagree with the parts. Simpson's paradox is easiest to grasp through an example. In the classic "Berkeley gender bias case" *(1)*, the University of California at Berkeley was sued for gender bias against women applicants based on the aggregate 1973 admission figures (44% men admitted vs. 35% women)—an observational dataset. The much higher male figure appeared to be damning. However, when individually looking at the men *vs*. women admission rate for each department, the rate was in fact similar for both sexes (and even in favour of women in most departments). The lower overall acceptance rate for women was not due to gender bias, but to the tendency of women to apply to more competitive departments, which have a lower admission rate in general. Thus, the association between gender and admission rate in the aggregate data could almost entirely be explained through strong association of these two variables with a third, confounding variable, the department. When controlling for the confounder, the association between the two first variables dramatically changes. This type of phenomenon is referred to as Simpson's paradox.

Because of the inherent heterogeneity of GO data, Simpson's paradox can manifest itself in GO analyses. This illustrates the importance of recognizing and controlling for potential biases and confounders.

## 1.2 The inherent incompleteness of the Gene Ontology (Open World Assumption)

The Gene Ontology is a representation of the current state of knowledge; thus, it is very dynamic. The ontology itself is constantly being improved to more accurately represent biology across all organisms. The ontology is augmented as new discoveries are made. At the same time, the creation of new annotations occurs at a rapid pace, aiming to keep up with published work. Despite these efforts, the information contained in the GO database, that is, the ontology and the association of ontology terms with genes and gene products, is necessarily incomplete. Thus, absence of evidence of function does not imply absence of function[1]. This is referred to as the Open World Assumption *(2, 3)*.

Associations between genes/gene products and GO terms ("annotations") are made via various methods: some manual, some automated based on the presence of protein domains or because they belong to certain protein families *(4)*. Annotations can also be transferred to orthologs by manual processes *(5)*, or automatically *(e.g. 6, 7, reviewed in 8)*. There are currently over 210 million annotations in the GO database. Despite these massive efforts to provide the widest possible coverage of gene products annotated, users should not expect each gene product to be annotated.

---

[1] Proteins whose function are uncharacterized are annotated to the root of the ontology, which formally means "this protein has *some* molecular_function/bp/cc, but a more specific assertion cannot be made". This annotation is associated with the evidence code "No biological Data available" (ND). The absence of annotation indicates that no curator has reviewed the literature for this gene product.

A further challenge is that the incompleteness in the GO is very uneven. Interestingly, the more comprehensively annotated parts of the GO can also pose challenges, presenting users with seemingly contradictory information (see section 3.2).

The inherent incompleteness of GO creates problems in the evaluation of computational methods. For instance, overlooking the Open World Assumption can lead to inflated false positive rates in the assessment of gene function prediction tools **(3)**. However, there are ways of coping with this uncertainty. For instance, it is possible to gauge the effect of incomplete annotations on conclusions by thinning annotations **(9)**, or analysing successive, increasingly complete database releases **(10, 11)**.

## 2. Gene Ontology structure

One potential source of bias is that not all parts of the GO have the same level of details. This has a strong implication on measuring the similarity of GO annotations (Chap XX). For instance, sister terms (terms directly attached to a common parent term) can be semantically very similar or very different in different parts of the GO structure, which has been called the "shallow annotation problem" **(e.g. 12, 13)**. This problem can partly be mitigated by the use of information-theoretic measures of similarity, instead of merely counting the number of edges separating terms, at the expense of requiring a considerable number of relevant annotations from which the frequency of co-occurrence of terms can be estimated (more details in Chap. [cross-reference to Sebastian Bauer's chapter]).

### *2.1 Understanding relationships between ontological concepts*

The GO is structured as a graph, and one pitfall of using the GO is to ignore this structure. Recall that each term is linked to other terms via different relationships (x-ref to Hasting's Ontology primer and to Gaudet et al's GO primer). These relationships need to be taken into account when using GO for data analysis.

Some relationships, such as "is a" and "part of", are *transitive*, which means that any protein annotated to a specific term is also implicitly annotated to all of its parents[2]. An illustration of this is a "serine/threonine protein kinase activity": it is a child of "protein kinase activity" with the relationship "is a". The transitivity of the relation means that the association between the protein and the term "serine/threonine protein kinase activity" and all its parents has the same meaning: the protein associated with "serine/threonine protein kinase activity" has this function, and it also has the more general function "protein kinase."

On the other hand, relations such as "regulates" are *non-transitive*. This implies that the semantics of the association of a gene to a GO term is not the same for its parent: if A is part of B, and B regulates C, we cannot make any inferences about the relationship between C and A. The same is true for positive and negative regulation. To illustrate, if we follow the term "peptidase inhibitor activity" (GO:0030414) to its parents, one of the terms encountered is "proteolysis" via a combination of "is a", "part of", and "regulates" relations. However a "peptidase inhibitor activity" does not *mediate* proteolysis; quite the contrary (Fig. 1). Thus, any logical reasoning on the ontology should take transitivity into account.

The relation "*has part*" is the inverse of "*part of*", and connects terms in the opposite direction. Because of this, it generates cycles in the ontology. The relation "*occurs in"* connects molecular function terms to the cellular components in which they occur. Thus, taking these relationships into account, it is possible to deduce additional cellular component annotations from molecular function annotations, without requiring additional experimental or computational evidence.

---

[2] With the exception of "NOT" annotations, for which the transitivity applies to *children* terms, not *parents* (see also Sect 3.2).

It important to know that there are three version of the GO ontology available: GO-basic, GO, and GO-plus[3]. Only the GO-basic file is completely acyclic. Therefore, applications requiring the traversal of the ontology graph usually assume that the graph is acyclic; hence, the GO-basic file should be used. The different GO ontology files are discussed in more details in Chap. XX (citation to Chap by Munoz-Torrez and Carbon).

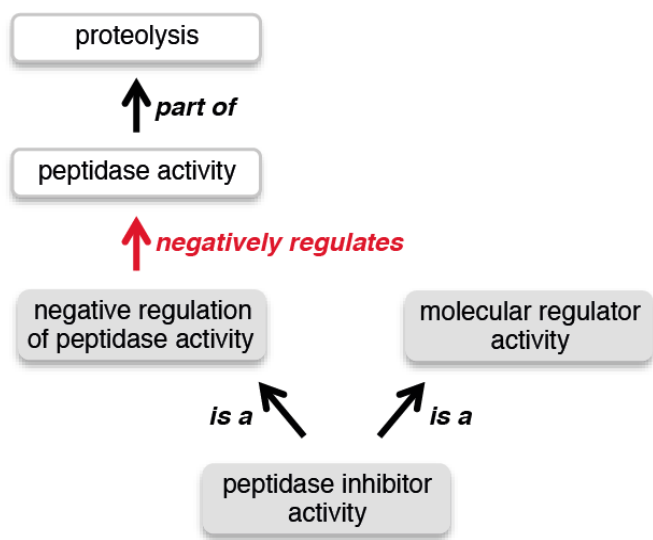

Figure 1. Example of transitive (black arrows) and non-transitive (red arrow) relationships between classes. A protein annotated to "peptidase inhibitor activity" term does not imply it has a role in "proteolysis", since the link is broken by the non-transitive relation *negatively regulates*.

### *2.2 Inter-ontology links and their impact on GO enrichment analyses*

The "part of" relation, when linking terms across the different *aspects* of the Gene Ontology (molecular function to biological process, or biological process to cellular component, for instance), triggers an annotation to the second term, using the same evidence code and the same reference, but "GOC" as the source of the annotation ("field 15 of the annotation file, see GO primer chapter [x-ref] for a description of the contents of the annotation file). For example, a DNA ligase activity annotation will automatically trigger an annotation to the biological process DNA ligation. The advantage of having these annotations inferred directly from the ontology is that it increases the annotation coverage by making annotations that may have been overlooked by the annotator when making the primary annotation. However, these inter-ontology links trigger a large number of annotations: there are currently 12 million annotations to 7 million proteins in the GO database. Changes in the structure of these links (as any change in the ontology), can potentially have a large impact on the annotation set. Indeed, Huntley *et al.* **(14)** reported that in November 2011, there was a decrease of ~2,500 manually and automatically assigned annotations to the term "transcription, DNA-dependent" (GO:0006351) due to the removal of an inter-ontology link between this term and the Molecular Function term "sequence-specific DNA binding transcription factor activity" (GO:0003700). Fig. 2 shows the strong and sudden variation in the number of annotations with term "ATPase activity" (GO:0016887).

---

[3] http://geneontology.org/page/download-ontology

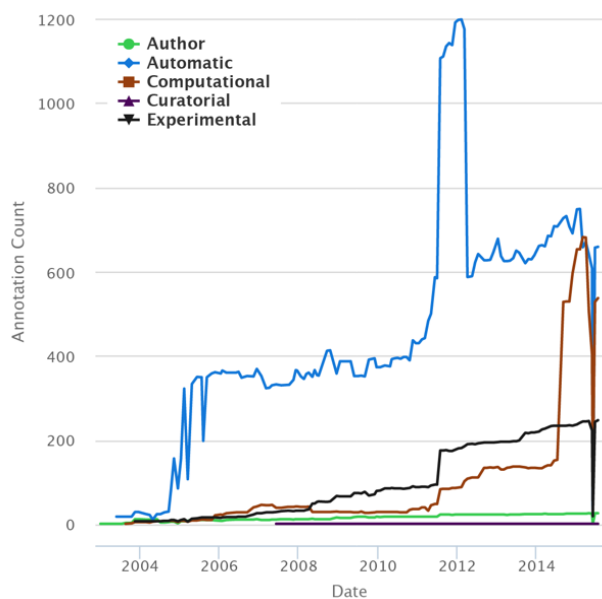

Fig 2. Strong and sudden variation in the number of annotations with the GO term "ATPase activity" (GO:0016887) over time. Such changes can heavily affect the estimation of the background distribution in enrichment analyses. To minimise this problem, use an up-to-date version of the ontology/annotations and ensure that conclusions drawn hold across recent releases. Data and plot obtained from GOTrack (http://www.chibi.ubc.ca/gotrack).

Such large changes in GO annotations can affect GO enrichment analyses, which are sensitive to the choice of background distribution *(Chap. S. Bauer; 15)*. For instance, Clarke and al. *(16)* have shown that changes in annotations contribute significantly to changes in over-represented terms in GO analysis. To mitigate this problem, researchers should analyse their datasets using the most up-to-date version of the ontology and annotations, and ensure that the conclusions they draw hold across multiple recent releases. At the time of the writing of this chapter, DAVID, a popular GO analysis tool, had not been updated since 2009 (http://david.abcc.ncifcrf.gov/forum/viewtopic.php?f=10&t=807). Enrichment analyses performed with it may thus identify terms whose distribution has substantially changed irrespective of the analysis of interest. The Gene Ontology Consortium now links to the PantherDB GO analysis service (http://amigo.geneontology.org/rte) *(17)*. This tool uses the most current version of the ontology and the annotations. Regardless of the tool used, researchers should disclose the ontology and annotation database releases used in their analyses.

## 3. Gene Ontology annotations

Having discussed common pitfalls associated with the ontology structure, we now turn our attention to annotations. Understanding how annotations are done is essential to correctly interpreting the data. In particular, the information provided for each GO annotation extends beyond the mere association of a term with a protein (reference to GO primer chapter). The full extent of this rich information, aimed to more precisely reflect the biology within the GO framework, is often overlooked.

## 3.1 Modification of annotation meaning by qualifiers

The Gene Ontology uses three qualifiers that modify the meaning of association between a gene-product and a Gene Ontology term: These are "NOT", "contributes to", and "co-localizes with" (see documentation at http://geneontology.org/page/go-qualifiers).

The "contributes to" qualifier is used to capture the molecular function of complexes when the activity is distributed over several subunits. However, in some cases the usage of the qualifier is more permissive, and all subunits of a complex are annotated to the same molecular function even if they do not make a direct contribution to that activity. For example, the rat G2/mitotic-specific cyclin-B1 CCNB1 is annotated as contributing to histone kinase activity, based on data in **(18)**, although it has only been shown to *regulate* the kinase activity of CDK1. Finding a cyclin annotated as having protein kinase activity may be unintuitive to users who fail to consider the "contributes to" qualifier.

The "co-localizes with" qualifier is used with two very different meanings: it first means that a protein is transiently or peripherally associated with an organelle or complex, while the second use is for cases where the resolution of an assay is not accurate enough to say that the gene product is a *bona fide* component member. Unfortunately, it is currently not possible to know which of the two meanings is meant in any given annotation.

## 3.2 Negative and contradictory results

The "NOT" qualifier is the one with the most impact, since it means that there is evidence that a gene product does *not* have a certain function. The "NOT" qualifier is mostly used when a specific function may be expected, but has shown to be missing, either based on closer review of the protein's primary sequence (e.g. loss of an active site residue) or because it cannot be experimentally detected using standard assays.

The existence of negative annotations can also lead to apparent contradictions. For instance, protein ARR2 in *Arabidopsis thaliana* is associated with "response to ethylene" (GO:0009723) both positively on the basis of a paper by Hass *et al.* **(19)** and negatively based on a paper by Mason *et al.* **(20)**. The latter discusses this contradiction as follows:

> *"Hass et al. (2004) reported a reduction in the ethylene sensitivity of seedlings containing an* arr2 *loss-of-function mutation. By contrast, we observed no significant difference from the wild type in the seedling ethylene response when we tested three independent* arr2 *insertion mutants, including the same mutant examined by Hass et al. (2004). This difference in results could arise from differences in growth conditions, for, unlike Hass et al. (2004), we used a medium containing Murashige and Skoog (MS) salts and inhibitors of ethylene biosynthesis."*

Thus, in this case, the contradiction in the GO is a reflection of the primary literature. As Mason *et al.* note, this is not necessarily reflective of a mistake, as there can be differences in activity across space (tissue, subcellular localisation) and time (due to regulation), with some of these details not fully captured in the experiment or in its representation in the GO.

A NOT annotation may also be assigned to a protein that does not have an activity typical of its homologs, for instance the STRADA pseudokinase (UniProtKB:Q7RTN6); STRADA adopts a closed conformation typical of active protein kinases and binds substrates, promoting a conformational change in the substrate, which is then phosphorylated by a "true" protein kinase, STK11 **(21)**. In this case, the "NOT" annotation is created to alert the user to the fact that although the sequence suggests that the protein has a certain activity, experimental evidence shows otherwise.

In contrast to positive annotations, "NOT" annotations propagate to children in the ontology graph and not to parents. To illustrate, a protein associated with a negative annotation to "protein kinase activity" is not a tyrosine protein kinase either, a more specific term.

### *3.3 Annotation extensions*

As also described in Chapter XX (reference to Lovering & Huntley's chapter), the Gene Ontology has recently introduced a mechanism, the "annotation extensions", by which contextual information can be provided to increase the expressivity of the annotations *(22)*. Until recently, annotations had consisted of an association between a gene product and a term from one of the three ontologies comprising the GO. With this new knowledge representation model, additional information about the context of a GO term such as the target gene or the location of a molecular function may be provided.

Common uses are to provide data regarding the location of the activity/process in which a protein or gene product participates. For example, the role of Mouse opsin-4 (MGI:1353425) in rhodopsin mediated signaling pathway is biologically relevant in retinal ganglion cells. Annotation extensions also allow capture of dynamic subcellular localization, such as the *S. pombe* bir1 protein (SPCC962.02c), which localizes to the spindle specifically during the mitotic anaphase. The annotation extensions can also be used to capture substrates of enzymes, which used to be outside the scope of GO.

The annotation extension data is available in the AmiGO *(23)* and QuickGO *(24)* browsers, as well as in the annotation files compliant with the GAF2.0 format (http://geneontology.org/page/go-annotation-file-gaf-format-20). However, because annotation extensions are relatively new, guidelines are still being developed, and some uses are inconsistent across different databases. Furthermore, most tools have yet to take this information into account.

In effect, extensions of an annotation create a "virtual" GO class that can be composed of more than one "actual" GO class, and can be traced up through multiple parent lineages. Thus, just as with inter-ontology links, accounting for annotation extensions can result in a substantial inflation in the number of annotations, which needs to be appropriately accounted for in enrichment analyses and other statistical analyses that require precise specification of GO term background distribution.

### *3.4 Biases associated with particular evidence codes*

Annotations are backed by different types of experiments or analyses categorised according to evidence codes (x-ref to GO Primer chapter Gaudet et al.). Different types of experiments provide varying degrees of precision and confidence with respect to the conclusions that can be derived from them. For most experiment types, it is not possible to provide a quantitative measure of confidence. Evidence codes are informative but cannot directly be used to exclude low-confidence data.[4] Nonetheless, the different evidence codes are prone to specific biases.

**Direct evidence.** Taking these caveats into account, the evidence code inferred from direct assay (abbreviated as IDA in the annotation files) provides the most reliable evidence with respect to the how directly a protein has been implicated in a given function, as it names implies.

**Mutant phenotype evidence.** Mutants are extremely useful to implicate genes products in pathways and processes; however exactly how the gene product is implicated in the process/function annotated is difficult to assess using phenotypic data because such data are inherently derivative. Therefore, associations between gene products and GO terms based on mutant phenotypes (abbreviated as IMP in the annotation files) may be weak. The same caveat applies to annotations derived from mutations in *multiple* genes, indicated by evidence code "inferred from genetic interaction" (IGI).

---

[4] An evidence confidence ontology has been proposed by Bastien *at al. (25)* but has yet to be adopted by the GO project.

***Physical interactions.*** Evidence based on physical interactions (IPI; mostly protein-protein interactions) is comparable in confidence to a direct assay for protein binding annotations or for cellular components; however for molecular functions and biological processes, the evidence is of the type "guilt by association" and is of low confidence. Inferences based on expression patterns (IEP) are typically of low confidence. The presence of a protein in a specific subcellular localization, at a specific developmental stage, or associated with a protein or a protein complex can provide a hint to uncover a protein's role in the absence of other evidence, but without more direct evidence that information is very weak.

***High-throughput experiments.*** Schnoes *et al.* ***(26)*** reported that annotations deriving from high-throughput experiments tend to consist of high-level GO terms, and tend to represent a limited number of functions. This artificially decreases the information content of these terms, since they are frequently annotated, and artificially decreased information content affects similarity analyses. This potentially has a large impact, since a significant fraction of the annotations in the GO database are derived from these types of analyses (as much as 25%, according to Schnoes *et al.*, who used the operational definition of a high throughput paper as one in which over 100 proteins were annotated). The GO does not currently record whether particular experimental annotations may be derived from high-throughput methods, but this may change in the future.

***Biases from automatic annotation methods.*** The GO association file, containing the annotations, has information regarding the method used to assign electronic annotations. The annotations can be assigned by a large number of different methods. Examples include domain functions, as assigned for example by InterPro, by Enzyme Commission numbers being associated with an entry, by BLAST, by orthology assignment, etc. Note that this information is not provided as an evidence code, but as a "reference code". The list of methods and their associated reference code is available at http://www.geneontology.org/cgi-bin/references.cgi. The large number of electronic annotations can also make them have a disproportionate impact on the results. Most analysis tools allow for the inclusion or exclusion of electronic annotations, but not at the more fine-grained level of the particular method. It is nevertheless possible to use the combination of evidence code plus reference (available at: http://www.geneontology.org/cgi-bin/references.cgi) to automatically deepen the evidence type, see https://raw.githubusercontent.com/evidenceontology/evidenceontology/master/gaf-eco-mapping.txt).

Note that a gene or gene product can have multiple annotations to the same term but with different evidence. This can provide corroborating information on particular genes, but may also require appropriate normalisation in statistical analyses of term frequency, as the frequency of terms that can be determined through multiple types of experiments may be artificially inflated. Furthermore, because different experiments can vary in their specificity—thus resulting in annotations at different levels of granularity for basically the same function—this redundancy only becomes conspicuous when the transitivity of the ontology structure is appropriately taken into account.

For more discussion on evidence codes, and their use in quality control pipelines, refer to Chap XX (Chibucos et al.).

## 3.5 Differences among species

There can be substantial differences in the nature and extent of GO annotations across different species. For instance, zebrafish is heavily studied in terms of developmental biology and embryogenesis while the rat is the standard model for toxicology. These differences are reflected in the frequency of GO terms across species, which can vary considerably across species ***(27)***. This has important implications on enrichment analyses and other statistical analyses requiring a background distribution of GO annotations. For instance, consider an experiment trying to establish the biological processes associated with a particular zebrafish protein by identifying its interaction partners and performing an enrichment analysis on them. If we naively use the entire database as background, the interaction partners might appear to be enriched in developmental genes simply because this class is over-represented in general in zebrafish. Instead, one should use zebrafish gene-related annotations only as background *(15)*.

### 3.6 Authorship bias

Other biases are less obvious but can nevertheless be strong and thus have a high potential to mislead. Recently, sets of annotations derived from the same scientific article we shown to be on average much more similar than annotations derived from different papers *(Fig. 3; 27)*. Unaware of this, Nehrt *et al.* compared the functional similarity of orthologs (genes related through speciation) across different species and paralogs (genes related through duplication) within the same species, and observed a much higher level of functional conservation among the latter *(28)*. However, this difference was almost entirely due to the fact that the GO functional annotations of same-species paralogs are ~50 times more likely to be derived from the same paper than orthologs; when controlling for the authorship, the difference in functional similarity between same-species paralogs and orthologs entirely vanished and even became in favour of orthologs *(27)*.

Note that the difference is smaller but remains significant if we compare annotations established from different papers, but with at least one author in common, with annotations from different articles with no author in common.

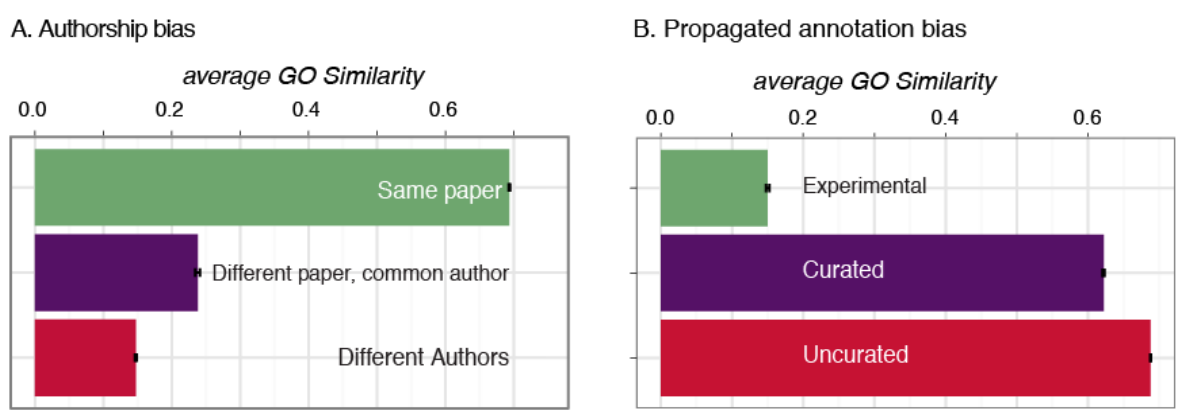

*Fig 3. (a) Average GO annotation similarity (using the measure of Schlicker et al. **(29)** between homologous genes, considering experimental annotations partitioned according to the provenance; (b) Average GO annotation similarity between homologous genes, partitioned according to their GO annotation evidence tags (Experimental: evidence code EXP and subcategories; Uncurated: evidence code IEA; Curated: all other evidence codes). Figure adapted from **(27)**.*

### 3.7 Annotator bias

Just as systematic differences among investigators can lead to the authorship bias, systematic differences in the way GO curators capture this information can lead to annotator bias. These annotator biases can in part be attributed to different annotation focus, but also to different interpretation or application of the GO annotation guidelines (http://geneontology.org/page/go-annotation-policies).

UniProt provides annotations for all species, which allows us to assess the effect of annotator (or database) bias. If we compare UniProt annotations for mouse proteins with those done by the Mouse Genome Informatics group (MGI), we see that comparable fractions of proteins are annotated using the different experimental evidence codes, with mutant phenotypes being the most widely used (78% of experimental annotations in MGI, versus 63% in UniProt, followed by direct assays (20% of annotations in MGI and 32% in UniProt).

However when we look at which GO terms are annotated based on phenotypes (IMP and IGI) by the two groups, we notice a large difference in the terms annotated. The top term annotated by MGI supported by the IMP evidence code is "in utero embryonic development", with 1170 annotations to 1020 proteins. UniProt has only 4 annotations for this term. On the other hand, UniProt has as one of its top-annotated classes "regulation of circadian rhythm", for 49 annotations to 38 proteins; 96 annotations for 69 proteins if we also include annotations to more specific, descendant terms. MGI on the other hand, only has 18 annotations for 19 proteins. This indicates that the annotations provided by different groups are biased towards specific aspects, and are not a uniform representation of the biology of all gene products in a species.

### *3.8 Propagation bias*

Another strong and perhaps surprising bias lies in the very different average GO similarity between electronic annotations compared with between experimental annotations. Indeed, if we consider homologous genes, their similarity in terms of electronic annotations tend to be much higher than in terms of experimental annotations, with curated annotations lying in-between **(27; Fig. 3)**. A likely explanation for this phenomenon is that electronic annotations are typically obtained by by inferring annotations among homologous sequences, a process that can only increase the average functional similarity of homologs.

Because of this homology inference bias, one must exercise caution when drawing conclusions from sets of genes whose annotations might have different proportions of experimental vs. electronic annotations. For instance, this would be the case when comparing annotations from model organisms with those from non-model organisms (the latter being likely to consist mostly of electronic annotations obtained through propagation).

More subtly, because function conservation is generally believed to correlate with sequence similarity, many computational methods preferentially infer function among phylogenetically close homologs. This bias can thus confound analyses attempting to gauge the conservation of gene function across different levels of species divergence.

### *3.9 Imbalance between positive and negative annotations*

As discussed above, both our knowledge of gene function and its representation in the GO remain very incomplete. We have already discussed the pitfalls of ignoring this fact altogether (closed vs. open world assumption), or assuming similar term frequencies across species. But the extent of missing data varies along other dimensions as well: for example it can depend on how easy it is to experimentally establish a particular function and how interesting the potential function might be. The problem is particularly acute in the case of negative annotations, because they can be even more difficult to establish than their positive counterparts (e.g. a negative result can also be due to inadequate experimental conditions, differences in spatio-temporal regulation, etc.) *and* they are often perceived as being less useful, and certainly less publishable. As a result, currently less than 1% of all experimental annotations are negative ones in UniProt-GOA **(30)**. This imbalance causes problems with training of machine learning algorithms *(31)*. Rider *et al. (32)* investigated the reliability of typical machine learning evaluation metrics (area under the "receiver operating characteristic" (ROC) curve, area under the precision-recall curve) under different levels of missing negative annotations and concluded that this bias could strongly affect the ranking obtained from the different metrics. Though this particular study adopted a closed world assumption, the effect of a varying proportion of negative annotations is likely to be even greater under the open world assumption.

## 4 Getting help

This chapter provides a broad overview of some of the pitfalls associated with GO-based analysis. Table 1 summarizes the most important pitfalls users encounter using GO.

Table 1: Main pitfalls or biases discussed in the chapter and their remedies.

| Pitfall or bias | Remedy |
| --- | --- |
| Wrongly assume that absence of annotation implies absence of function. | Account for the fact that both ontology and annotations are necessary incomplete, for instance by assessing the impact of incompleteness on one's analyses and findings. |
| Not all directed edges in the ontology structure have the same meaning: depending on their type, the relationship they represent may or may not be transitive. | The transitivity of each type of relations must be taken into account when reasoning over the GO. "Is a" and "part of" are transitive, but "regulates" and "has part" are not. |
| To yield meaningful results, GO enrichment analyses require accurate specification of the background distribution, which can vary substantially across releases, species, etc. | Specify the actual background distribution used in the analysis of interest. Short of this, ensure that the enrichment analysis is performed on consistent database release and subsets of species, terms, etc. To test the robustness of results, consider repeating the analysis using several releases of GO ontology/annotation databases. Avoid tools that are not regularly updated. |
| Inter-ontology links and annotation extensions can result in large variations in the number of annotations. Furthermore, annotation extensions may not be consistently implemented, if at all, across analyses tools or workflows. | Keep track of database releases in analyses. If they are relevant, make sure that annotation extensions are implemented consistently. |
| Qualifiers such as "NOT" or "co-localizes with" are important parts of a gene annotation in that they fundamentally change the meaning of annotations. Because only a small minority of all annotations have qualifiers, such errors can easily go unnoticed. | Remember to take into account qualifiers. When using tools or software libraries, make sure that these take qualifiers into account as well. |
| Annotations are supported by different types of evidence (categorised by evidence codes). The annotations associated with each code vary in their scope, specificity, and number. These differences can confound some analyses. | Take evidence code into account. In statistical analyses, consider the distribution of annotations in terms of evidence codes, and, if needed, control for this potential confounder. |
| Different species tend to have very different | When performing statistical analyses or using |
| types of annotations. For instance, model species have many more experiment-based annotations. | information-theoretic similarity measures, use species-specific frequencies of GO term. |
| Experiment-based annotations derived from the same research article tend to be more similar than annotations derived from different articles. Similar trends hold for annotations derived from same versus different authors, and same versus different annotators. | Control for authorship bias in analyses that may have varying proportion of annotations stemming from the same article, lab, or annotation team. |
| Because annotations are preferentially propagated among closely related sequences, electronic annotations can confound analyses seeking to characterise relationships between evolution and function. | Restrict such analyses to experiment-based annotations. Avoid circularity. |
| There are many more positive annotations than negative annotations. As a result, standard accuracy measures used by machine learning methods may be misleading ("class imbalance problem"). | Consider false-positive and false-negative rates separately. Focus on subset of data for which the class imbalance problem is less pronounced. |

Users are advised to make use of a number of excellent resources provided by the GO consortium:
- The GO website http://geneontology.org
- The GO FAQ http://geneontology.org/faq-page
- The GO team are eager to help with with your problems: email go-help@geneontology.org
- The wider bioinformatics community can be consulted via sites like biostars – see the GO tag https://www.biostars.org/t/go/
- The GO community can be contacted on Twitter at @news4go

## 5 Conclusion

This chapter has surveyed some of the main pitfalls and biases of the Gene Ontology. The number of potential issues, summarised in Table 1, may seem daunting. Indeed, as discussed at the start of this chapter, there are some inherent risks in working with observational data. However, simple remedies are available for many of these (Table 1). By understanding the subtleties of the GO, controlling for known confounders, trying to identify unknown ones, and cautiously proceeding forward, users can make the most of the formidable resource that is the GO.

## Acknowledgements

We thank Natasha Glover, Rachel Huntley, Suzanna Lewis, Chris Mungall, and Paul Thomas for detailed and helpful feedback on the manuscript. PG acknowledges National Institutes of Health/National Human Genome Research Institute grant HG002273. CD acknowledges Swiss National Science Foundation grant 150654 and UK BBSRC grant BB/M015009/1.